\documentclass[grl]{agutex}

\usepackage{url}
\usepackage{graphicx}
\usepackage{lineno}
\setkeys{Gin}{draft=false}

\def\tmae{Terr. Magn. Atmos. Elec., }
\def\grl{Geophys. Res. Lett., }  
\def\jgr{J. Geophys. Res., }  
\def\ang{Ann. Geophys., } 
\def\adg{Ann. de G\'eophys., }  
\def\rgsp{Rev. Geophys. Space Phys., }  
\def\sol{Solar Phys., }  
\def\trse{Trans. Roy. Soc. Edinburgh, }
\def\ptrsl{Phil. Trans. Roy. Soc. London A, }
\def\eos{EOS, Trans. AGU, }
\newcommand{\hide}[1]{}

\authorrunninghead{SVALGAARD}

\titlerunninghead{SEMIANNUAL VARIATION}

\begin{document}

\title{Geomagnetic Semiannual Variation Is Not Overestimated and Is Not an Artifact of Systematic Solar Hemispheric Asymmetry}

\authors{L. Svalgaard\altaffilmark{1}}
\altaffiltext{1}{HEPL, Via Ortega, Stanford University, Stanford, CA 94305, USA. (leif@leif.org)}

\begin{abstract}
\cite{mur11} (MTL11) suggest that there is a 22-year variation in solar wind activity that coupled with the variation in heliographic latitude of the Earth during the year, gives rise to an apparent semiannual variation of geomagnetic activity in averages obtained over several solar cycles. They conclude that the observed semiannual variation is seriously overestimated and is largely an artifact of this inferred 22-year variation. We show: (1) that there is no systematically alternating annual variation of geomagnetic activity or of the solar driver, changing with the polarity of the solar polar fields, (2) that the universal time variation of geomagnetic activity at all times have the characteristic imprint of the equinoctial hypothesis rather than that of the axial hypothesis required by the suggestion of MTL11, and (3) that the semiannual variation is not an artifact, is not overestimated, and does not need revision.
\end{abstract}

\begin{article}

\section{Introduction}
\cite{mur11} noticed that the strongest geomagnetic activity during the short 16-year interval 1993-2008 occurred when the Earth was at southern heliographic latitudes in 1994 and at northern heliographic latitudes in 2003. These two observations are presented as evidence that there is an annual variation of solar wind speed at Earth, changing phase between the two cycles resulting in a predominance of high speed streams from the southern hemisphere during the late phase of cycle 22 and from the northern hemisphere during the late phase of cycle 23 (although the high geomagnetic activity in October 2003 originated from solar activity in the southern hemisphere), and that this alternation is characteristic of solar cycles in general, providing long-term predictability of activity. The phase of the purported effect is such that the Earth experiences the fastest solar wind when it is north of the solar equator during positive polarity epochs (magnetic field positive (outward) at the northern pole of the Sun) such as during the decline of sunspot cycle 22 from 1990-1996 and when it is south of the solar equator during negative polarity epochs such as the decline of sunspot cycle 23 (2000-2008).

MTL11 suggest a novel variation on the axial mechanism, namely that this inferred 22-year variation in solar wind activity, coupled with the variation in heliographic latitude of the Earth during the year, gives rise to an annual variation of geomagnetic activity with opposite phases between the  declining phases of the cycles, and that these two annual waves when averaged over several cycles result in an apparent semiannual variation. They conclude that the observed semiannual variation is seriously overestimated and is largely an artifact of a 22-year variation. We examine this claim using the full geomagnetic record extending back well into the 19th century and find it unjustified.

\section{The Semiannual and Universal Time Variations}
\begin{figure}
\centerline{\includegraphics[width=0.45\textwidth]{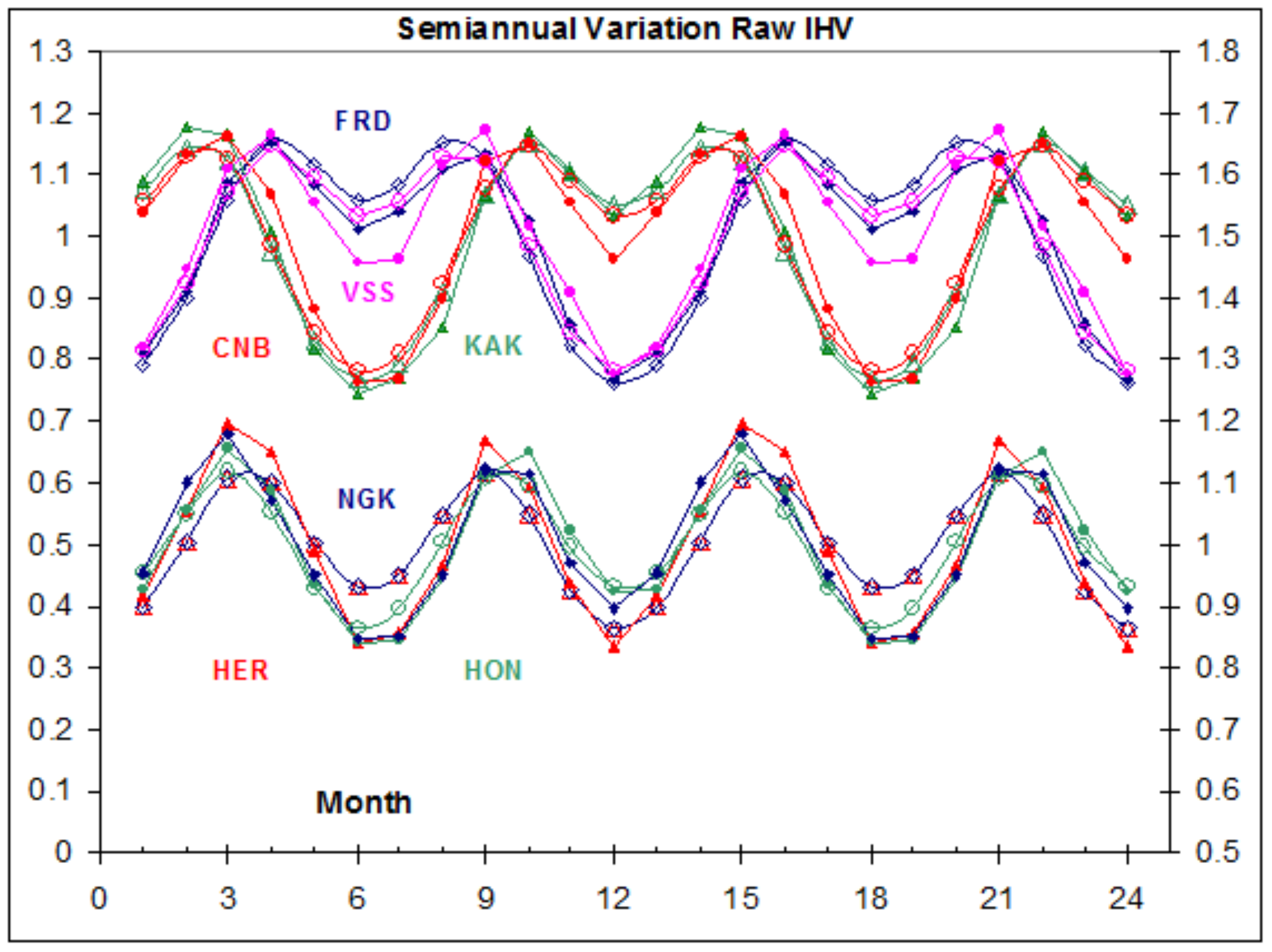}} 
\caption{Semiannual variation of Raw {\it IHV} for stations at different geographic longitudes; southern hemisphere stations: reddish color. Upper curves (left-hand scale) for stations with longitude near those of the geomagnetic poles. Lower curves (right-hand scale) for stations with longitudes $\sim$90\deg~away from the poles. Filled symbols show observations (roughly spanning the 20th century using predecessor stations as needed), while open symbols show the variation calculated from the Svalgaard Function, $S(\Psi)=1.175/(1+3cos^2(\Psi))^{2/3}$, \citep{obr02,sva77}. All values are normalized to the mean for the year and repeated for one year in the right-hand half of the Figure.
} \label{Obs-semiannual} 
\end{figure}

As MTL11 point out, the semiannual variation is a problem of long standing \citep{mai33, bro48, bar32, mci59, cli00}. Geomagnetic activity is a very complicated phenomenon. To reduce the complexity we commonly resort to study geomagnetic indices instead. Many indices exist (Ci, u, Dst, AE, IL, ap, aa, am, IHV, etc),  e.g. \cite{men11}. When Julius Bartels designed the geomagnetic activity index {\it Kp} \citep{bar39,bar49}, he found that the {\it K} indices for each station had such a strong dependence on local time that the very non-uniform longitudinal distribution of the {\it Kp} stations precluded any investigations of the expected  Universal Time (UT) of the activity as the angle, $\Psi$, between the direction to the Sun and the Earth's magnetic dipole varies both with day of year and with hour of universal time, thus introducing different semiannual variations depending on Universal Time. Thus it was necessary to remove the local time variation by applying normalization factors for each station for each month. The resulting {\it Kp} index should then have no UT variation at all. The {\it ap} index derived by linearizing {\it Kp} is therefore, by design, not suitable for a comprehensive study of the semiannual variation of geomagnetic activity. Figure~\ref{Obs-semiannual} shows how strongly the UT-variation modulates the semiannual variations at individual stations, using the `raw' IHV index for each station \citep{sva07}. {\it IHV} measures geomagnetic activity for an interval around local midnight where the contribution (or interference) from the diurnal $Sq$ variation is minimal. Calculating the curves using local {\it K} indices (converted to amplitudes), rather than {\it IHV} yields identical results, as expected, because both indices respond the same to the same physical processes. 

\begin{figure}
\centerline{\includegraphics[width=0.45\textwidth]{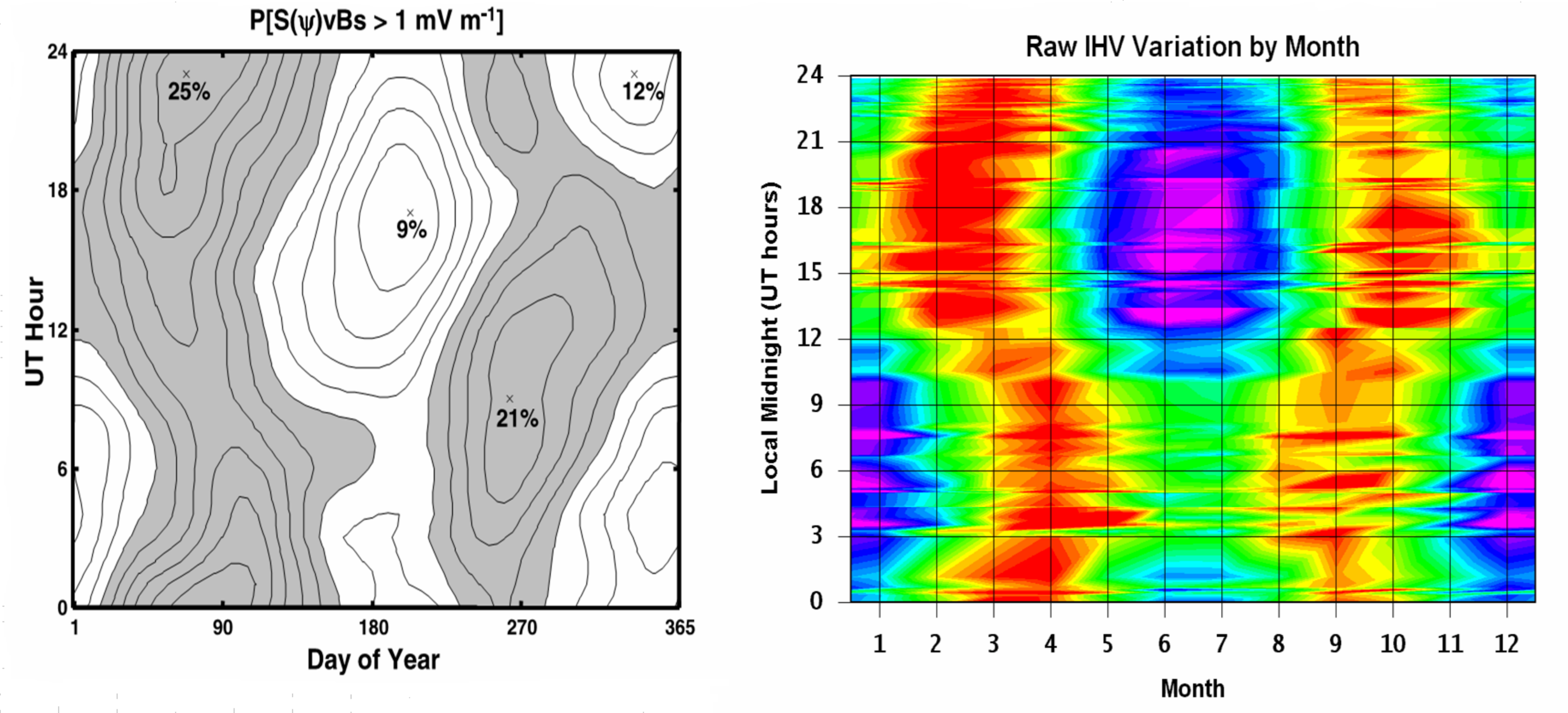}}
\caption{Left: Probability of ring current injection \citep{obr02}. Right: Variation of ``raw'' {\it IHV} \citep{sva07} for the $\approx$60 stations used to calculate {\it IHV}. Warm (red) colors signify maxima and cold (purple) colors minima. {\it IHV} measures geomagnetic activity for an interval around local midnight where the contribution (or interference) from the regular diurnal $Sq$ variation is minimal.
} \label{Dst-IHV} 
\end{figure}

\cite{may67,may70} utilized a more uniform distribution of geomagnetic observatories in both hemispheres to construct a true planetary activity index, {\it Km}, and its linear version {\it am}. Following \cite{sva77}, \cite{cli00} plotted the {\it am} index as 2D-contours as a function of month of year and of universal time, to reveal a characteristic `hourglass' structure. The probability of injection of energy into the ring current \citep{obr02} and the `raw' {\it IHV} index \citep{sva07} display, Figure~\ref{Dst-IHV}, the same characteristic hourglass structure, showing that the semiannual/UT variation is not just an artifact of the {\it am} index. New indices based on the same principles produce the same hourglass signature, e.g. Figure 3 of \cite{fin08}. Importantly, they show that for the last 50 years, a time when the 22-year solar wind-variability claimed by MTL11 to be particularly well-defined, geomagnetic indices show the clear imprint of the equinoctial hypothesis for the semiannual variation of geomagnetic activity, \citep{bar32,mci59,sva77,cli00}, rather than the absence of the UT-variation one would expect to see if the axial effect invoked by MTL11 were the dominant cause of the semiannual variation.

\begin{figure}
\centerline{\includegraphics[width=0.45\textwidth]{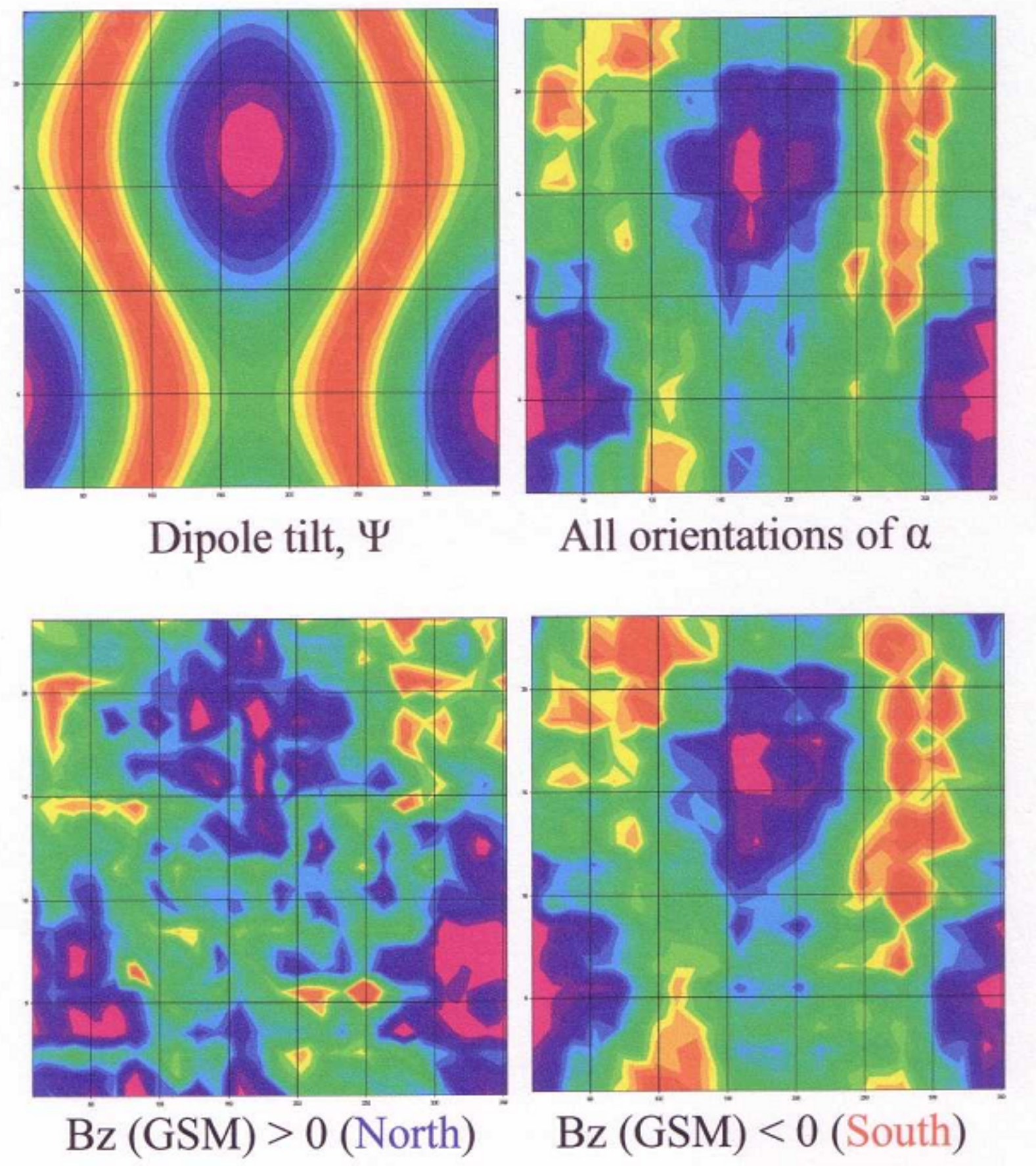}} 
\caption{The variation of {\it am} reduced for the influence of IMF strength and solar wind speed for the interval 1963-2003 for all times where we have simultaneous {\it am}, $B$, and $V$ data, from \cite{sva04}. The lower panel shows that the existence of the `hourglass' structure does not depend on the orientation of the IMF.
} \label{Psi-Am-N-S} 
\end{figure}

\begin{figure}
\centerline{\includegraphics[width=0.45\textwidth]{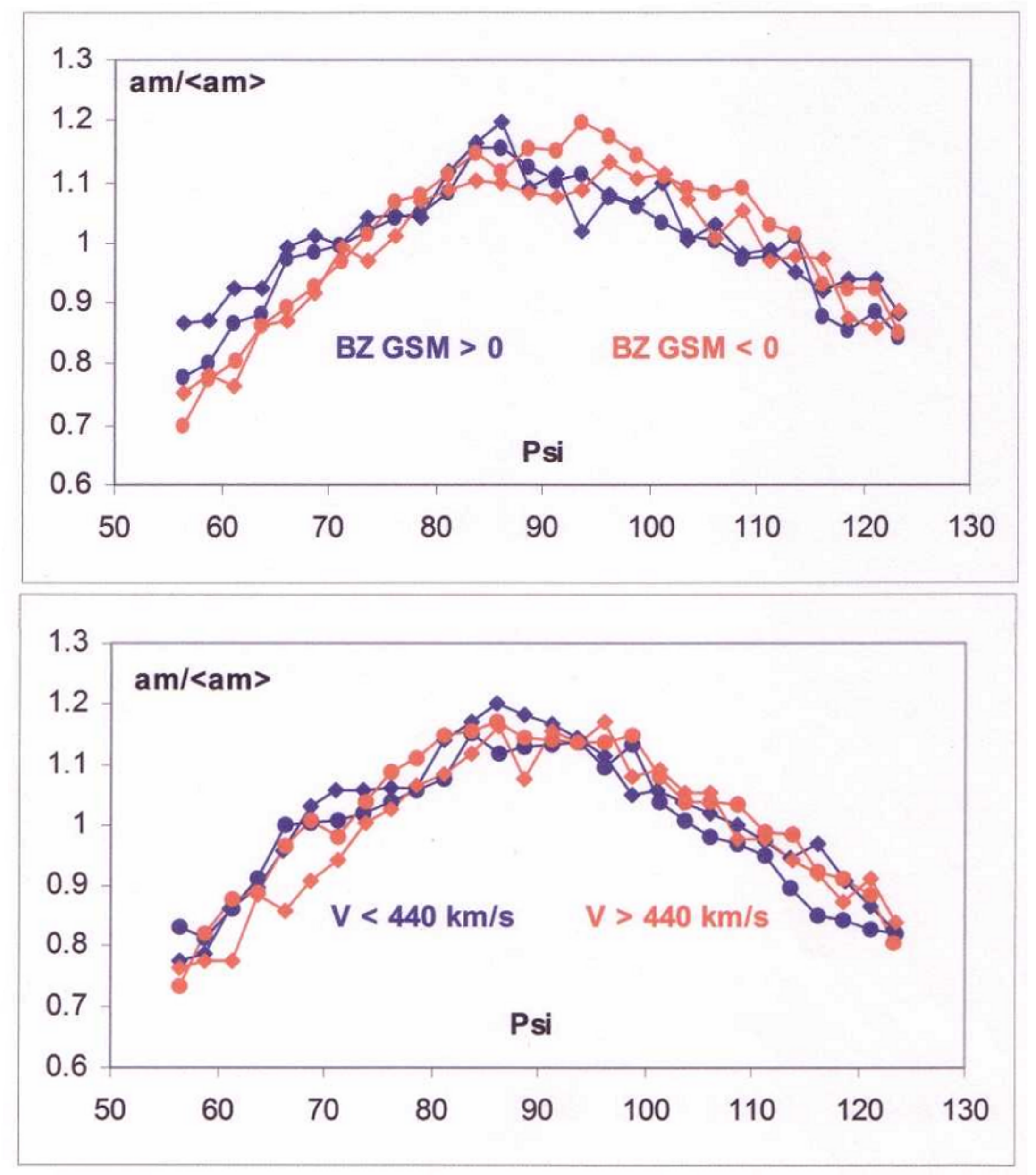}} 
\caption{Upper: Relative variation of {\it am} (i.e. divided by the mean value) as a function of dipole tilt angle, $\Psi$, separately for northward (GSM, blue) and southward IMF (red) for odd/even years (diamonds/dots) from \cite{sva04}. Lower: same but for low/high solar wind speed.
} \label{Am-Bz-V} 
\end{figure}

\section{Solar Wind Drivers}
\cite{sva77} (see also \cite{cro77} and \cite{mur82}) has shown that the {\it am} index can with good approximation be calculated as $am=k\ q(\alpha, DOY, UT) B V^2$ where $k$ is a scale factor, $q$ is a geometrical factor depending on the IMF clock angle ($\alpha$), on day of year ($DOY$), and on Universal time ($UT$), $B$ is the magnitude of the IMF strength, and $V$ is the solar wind speed, all at Earth. To separate to first order the geometric effects from the intrinsic variations of the solar wind $B$ and $V$, we can divide the geomagnetic activity index under investigation by the product $BV^2$. Figure~\ref{Psi-Am-N-S} shows the resulting hourglass (right-hand panel) and the variation of the dipole tilt angle. It is clear that the semiannual variation is closely regulated by the dipole tilt towards the solar wind, regardless of $B$ and $V$. In fact, the hourglass is found for both northward and southward IMF as shown in the lower panels.

Plotting {\it am} (normalized to its average) against the tilt angle, separately for $B$ and $V$ shows, Figure~\ref{Am-Bz-V}, that the pattern of semiannual/UT variation of activity with the dipole tilt is independent of the intrinsic values of these solar wind parameters. The semiannual/UT variation is a {\it permanent} feature of the interaction between the solar wind and the magnetosphere, its functional form independent of the direction and strength of the IMF and of the speed of the solar wind, that is: The dominant semiannual-UT variation is a {\it modulation} of activity generated by the impact of the solar wind, as has been known for decades \citep{may77,sva77}.  The notion, put forward in MTL11 and elsewhere \citep{kar06, lov09}, that the semiannual variation is `excessive', an artifact, due to uneven station distribution,  incomplete removal of the regular diurnal $Sq$ variation, or other deficiencies in deriving geomagnetic indices, is contradicted by the regularity of the combined semiannual-UT variation observed at stations of every longitude, for all solar wind conditions, for different indices, as well as for related geophysical phenomena \citep{mai33,bak99}.     

\section{Annual Variation of Solar Wind Parameters}
\begin{figure}
\centerline{\includegraphics[width=0.45\textwidth]{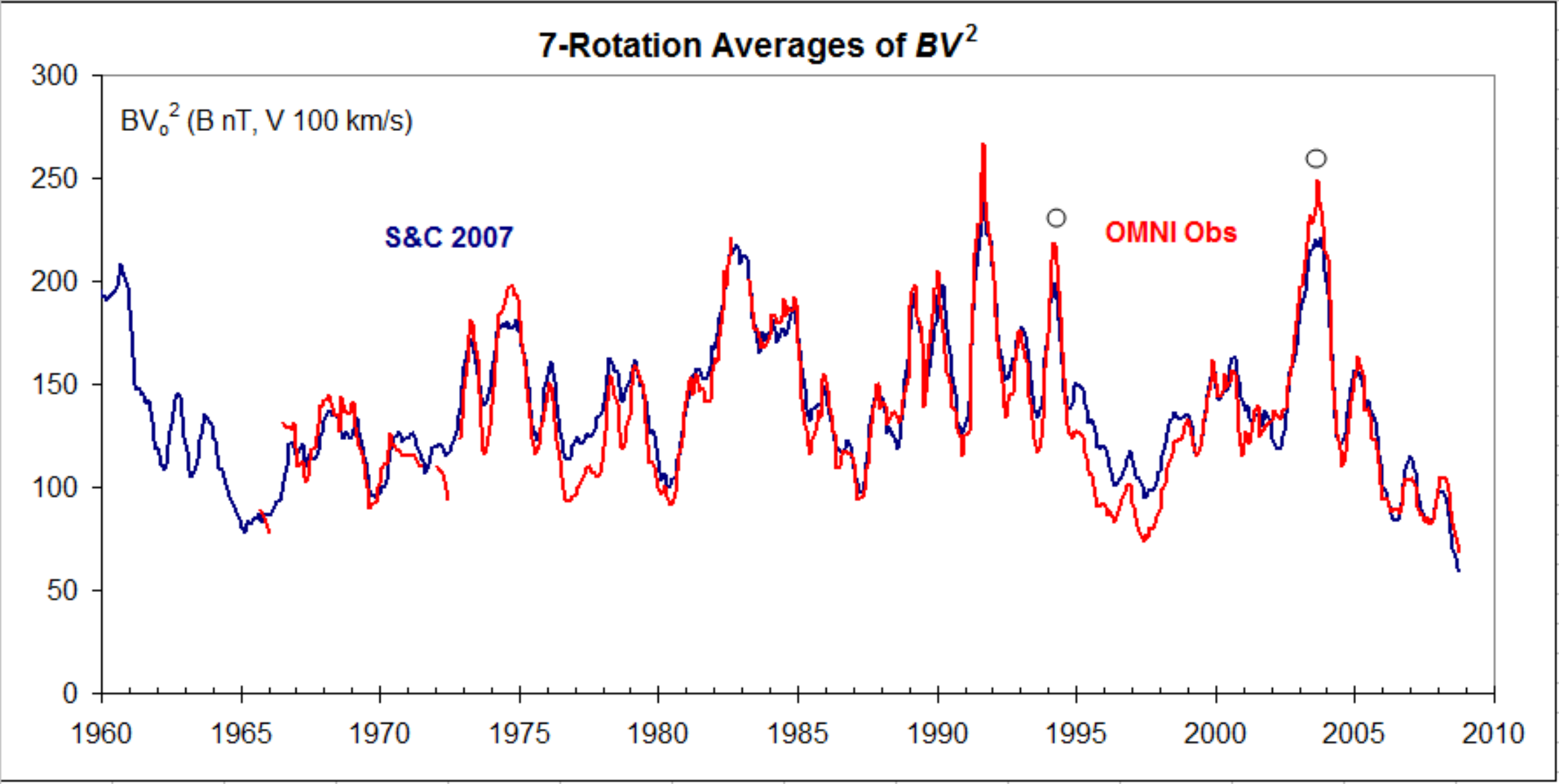}}
\caption{Seven-rotation running means of $B~V^2$ ($B$ in nT and $V$ in units of 100 km/s) calculated from the {\it IHV-IDV} indices, \citep{sva07,sva10} blue curve, and determined from in situ observed solar wind near the Earth (OMNI: \url{http://omniweb.gsfc.nasa.gov} red curve). Circles mark the high-speed streams in 1994 and 2003 that seemed so important to MTL11. Only rotations with better than 50\% coverage were included in the calculation. Note the 22-year variation \citep{cli04} in geomagnetic activity resulting in calculated values that are slightly too high during the rising phase of solar cycles 21 and 23.
} \label{BV2-recon-obs} 
\end{figure}

\begin{figure}
\centerline{\includegraphics[width=0.45\textwidth]{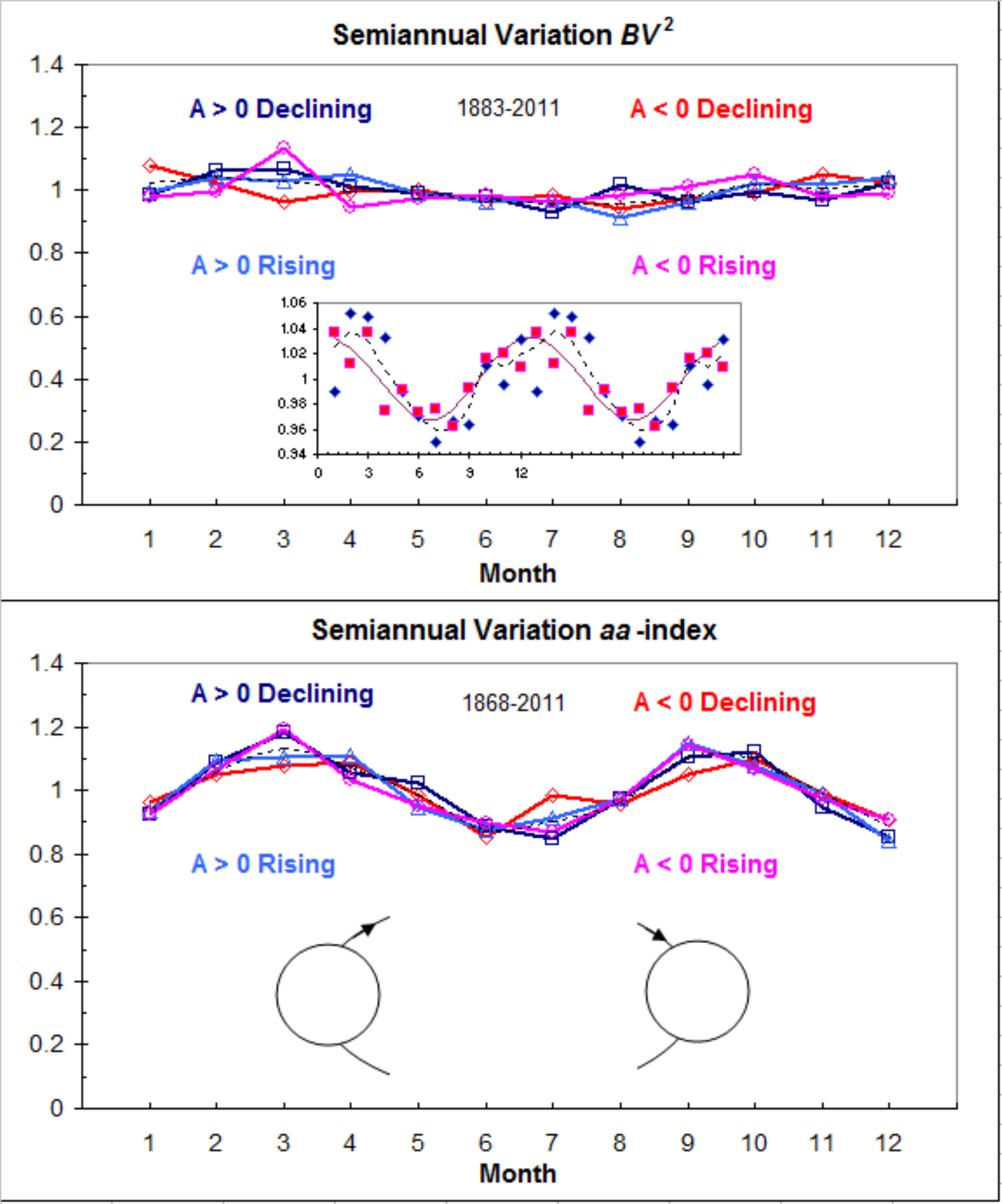}}
\caption{Upper: Variation of the solar wind driver $BV^2$ through the year, normalized to mean values. Bluish colors for North Pole positive ($A>0$; dark blue, squares: declining phase; light blue, triangles: rising phase), reddish colors for negative ($A<0$; red, diamonds: declining; pink, circles: rising). Dashed black curve: all data. Insert: variation of normalized $BV^2$ (blue symbols: $A>0$; red symbols: $A<0$; dashed line: all data; full purple curve: the inverse square of the solar distance in AU). Lower: Variation of {\it aa} index; same symbols and colors as for the upper plot.
} \label{BV2-aa} 
\end{figure}

\cite{sva07,sva10} show how both $B$ and $V$ can be determined from the geomagnetic record (the {\it IDV} and {\it IHV} indices). Figure~\ref{BV2-recon-obs} shows 7-rotation running averages of the product $B~V^2$ comparing the value reconstructed from geomagnetic activity directly with in situ observations, covering the whole of the space age. It is clear that the reconstructed values are a good representation of the physical reality when several rotations are averaged (for individual rotations the assumption of balanced northward and southward fields occasionally breaks down). MTL11 suggest that the phase of the annual variation of solar wind speed and geomagnetic activity changes systematically from one solar cycle to another, that the annual variation is largest in the declining phase of solar cycles, and that annual maxima are located in March during positive polarity periods and in September during negative polarity periods. As shown by \cite{sva07} there is a true, but small (5\%), second order, variation of geomagnetic activity caused by the variation of $B$ with distance to the Sun, as expected (their Figure A7 and the insert in Figure~\ref{BV2-aa}). This small variation, attesting to the validity of the reconstructed solar wind parameters, is, of course, independent of the polarities of the solar cycles, as the orbit of the Earth does not change with the cycles, and is not in phase with the variations claimed by MTL11. In addition, at any time, the solar wind can have enhancements of some duration, especially in $V$ as were the cases for the high-speed streams of 1994 and 2003, or in HMF-polarity anomalies as in 1954 and 1996, e.g. \cite{cli04}. 

The central thesis of MTL11's mechanism for the semiannual variation can be stated as follows: The observed solar drivers, the excitation parameters $B$ and $V$, do not vary randomly with respect to the Earth's orbit and their variation, when coupled with the axial effect, is of sufficient amplitude to be the dominant cause of the observed semiannual variation. We can put this thesis to a direct test using as our measures of the solar driver and of geomagnetic activity the long-term variation of $BV^2$ (directly observed since 1965 and reconstructed from {\it IDV} and {\it IHV} before that) and of the {\it aa} index (which is adequate for monthly values). We consider four subsets of the series: rising and declining phases of the cycle and for each of those, positive and negative polarity (of the northern pole), and then calculate the variations of $BV^2$ and of {\it aa} as a function of month of year. The data for this calculation is given in the electronic supplement. It is plain from the resulting Figure~\ref{BV2-aa} that during the declining phase there is no asymmetry in solar wind parameters nor in geomagnetic activity between the first half of the year and the second half, contrary to the claims of MTL11. For the rising phase, there is no asymmetry either. The claims thus fail this direct test. 

\begin{figure}
\centerline{\includegraphics[width=0.45\textwidth]{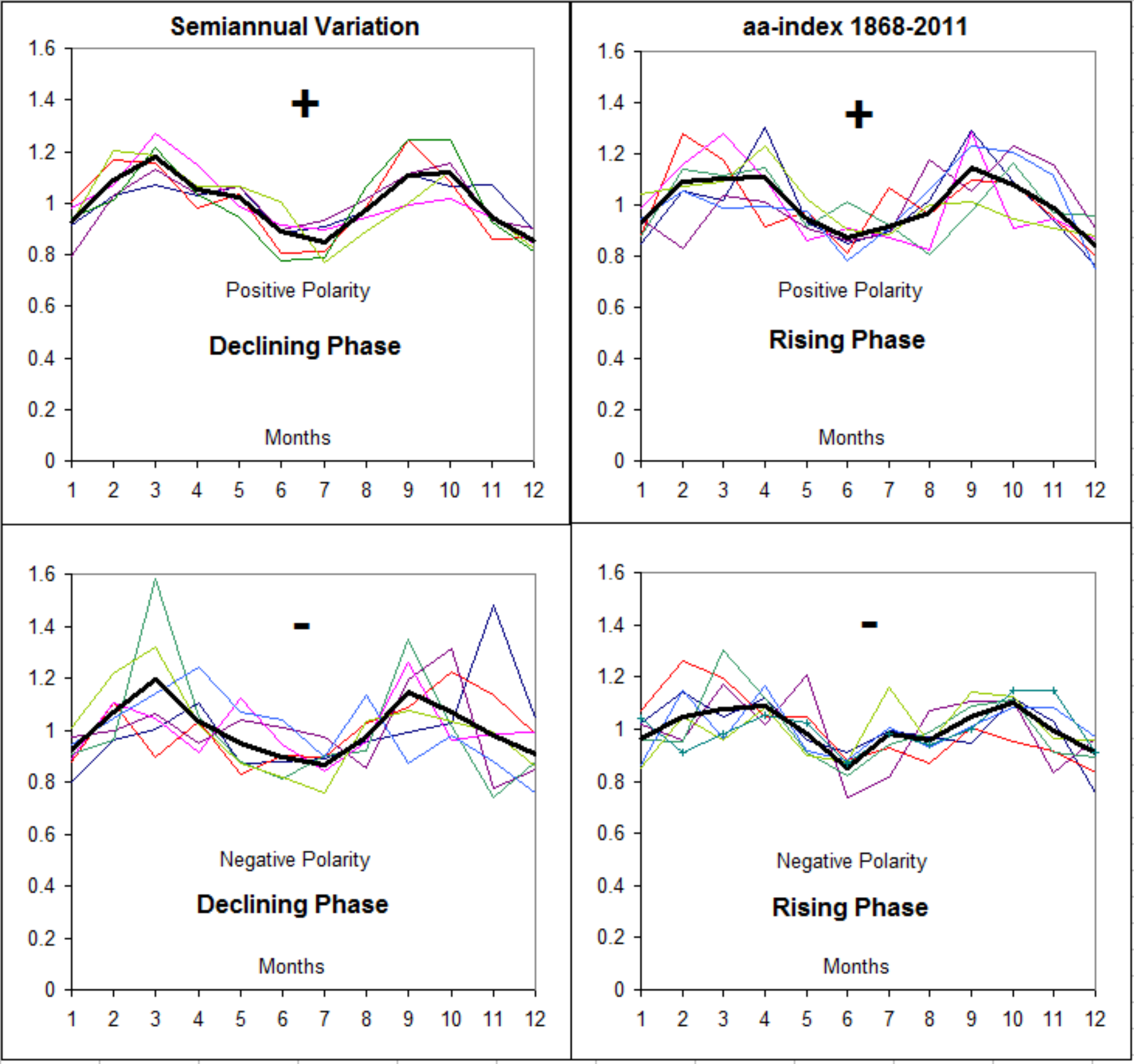}}
\caption{Variation of the {\it aa} index (1868-2011) separately for the declining and rising phases of each solar cycle and for each polarity, positive or negative. The thick, black curve shows the average variation, while thin, colored curves are for individual solar cycles. 
} \label{Aa-semiannual-details} 
\end{figure}

As average curves at times may conceal actual variation, we show in  Figure~\ref{Aa-semiannual-details} for individual cycles since 1868 that there is no systematic annual variation in the {\it aa} index, neither for the declining phase or the rising phase, nor for either polarity. To compare cycles, the values for each section of a cycle were normalized by the average for that section. 

\section{The 22-year Variation}
\cite{che66} discovered a 22-year variation in geomagnetic activity. The effect is explained \citep{rus75,sva77} by a combination of the Rosenberg-Coleman effect \citep{ros69,wil72} varying the dominant polarity of the heliospheric magnetic field and the Russell-McPherron effect \citep{rus73}, favoring enhanced geomagnetic activity when the solar and terrestrial magnetic dipoles have opposite directions \citep{cli04}. This is a purely geometrical effect, not dependent on any inherent hemispheric asymmetries in solar activity. \cite{ech04} showed that the Rosenberg-Coleman effect is only prominent from minimum through the rising phase of the solar cycle, so the 22-year cycle is confined to such times (the conclusion of \cite{ech04} that the 22-year cycle must have another explanation is probably too simplistic as they did not appreciate that the enhancement of geomagnetic activity is also confined to the same phase of the cycle as the Rosenberg-Coleman effect), as is also evident in Figure~\ref{BV2-recon-obs}. On rare occasions such as in 1954 and 1996 \citep{cli04} this effect (not tied to any inherent enhancement of solar activity, e.g. of solar wind speed) can at those times dominate the semiannual variation, thus making the minor, overall contribution to the average semiannual variation reported by \cite{sva02} and \cite{cli04}. This (real) 22-year cycle is very distinct from and has nothing to do with the variation of solar wind speed claimed by MTL11.  

\section{Conclusion}
MTL11 make two claims 1: that northern and southern solar hemispheres have a 22-year cycle in systematically different, and opposite, activity levels resulting in corresponding variations of the solar drivers of geomagnetic activity, primarily solar wind speed. When the Earth is north (south) of the solar equator, a more active northern (southern) hemisphere would result in a stronger and faster solar wind causing enhanced geomagnetic activity, and 2: that the well-established semiannual variation is largely an axial effect resulting from the above asymmetry, averaging two disparate annual variations offset by six months resulting in an artificial semiannual variation.

We show here that there are no oppositely organized annual variations in the solar driver of geomagnetic activity nor in the observed activity in step with the alternating cycle polarities, precluding long-term predictive capability. Finally, such a purely axial mechanism does not allow for the UT variation which is a well-established and integral part of the phenomenon. Consequently, the semiannual variation is not `seriously overestimated' and is not in need of revision.

\begin{acknowledgments}
I thank Ed Cliver for useful discussions and I acknowledge with appreciation the archiving efforts of the OMNI and {\it aa} datasets as well as the immense labor involved in observing the geomagnetic field through centuries.
\end{acknowledgments}

% REFERENCE LIST AND TEXT CITATIONS

\end{article}

\end{document}